\documentclass[namedreferences]{solarphysics}
\usepackage[hyperref,optionalrh,solaromanenum]{spr-sola-addons} 
\usepackage{graphicx}                    
\usepackage{color}                       
\usepackage{breakurl}                         

\newcommand{\araa}{   {\it Ann. Rev. Astron. \&\ Astrophys.}}
\newcommand{\aap}{    {\it Astron. Astrophys.}}

\newcommand{\aapr}{   {\it Astron. Astrophys. Rev.}}

\newcommand{\apj}{    {\it Astrophys. J.}}

\newcommand{\apo}{    {\it Appl. Opt.}}

\newcommand{\grl}{    {\it Geophys. Res. Lett.}}

\newcommand{\memsai}{ {\it Memorie della Società Astronomica Italiana}}

\newcommand{\solphys}{{\it Solar Phys.}}

\newcommand{\ssr}{    {\it Space Sci. Rev.}}
\newcommand{\caii}    {Ca {\sc ii} K~}
\newcommand{\exgr}   {\textit{e.g.}~}
\begin{document}

\begin{article}

\begin{opening}

\title{Correlation Between Sunspot Number and \mbox{\caii} Emission Index}

%
\author[addressref={aff1},corref,email={lbertello@nso.edu}]{\inits{L.}\fnm{Luca}~\lnm{Bertello}}
\author[addressref={aff2},email={apevtsov@nso.edu}]{\inits{A.A.}\fnm{Alexei}~\lnm{Pevtsov}}
\author[addressref={aff3},email={tlatov@mail.ru}]{\inits{A.G.}\fnm{Andrey}~\lnm{Tlatov}}
\author[addressref={aff4},email={jsingh@iiap.res.in}]{\inits{J.}\fnm{Jagdev}~\lnm{Singh}}

\address[id=aff1]{National Solar Observatory, Boulder, CO 80303, USA}
\address[id=aff2]{National Solar Observatory, Sunspot, NM 88349, USA}
\address[id=aff3]{Kislovodsk Solar Station of Pulkovo Observatory, Kislovodsk, Russia}
\address[id=aff4]{Indian Institute of Astrophysics, Koramangala, Bangalore, India}

%


\begin{abstract}
Long-term synoptic observations in the resonance line of \caii constitute a fundamental 
database for a variety of retrospective analyses of the state of the solar magnetism. Synoptic 
\caii observations began  
in late 1904
at the Kodaikanal Observatory, in India. 
In early 1970s, the National Solar Observatory (NSO) at Sacramento Peak (USA) started a new program of daily Sun-as-a-star 
observations in the \caii line. Today the NSO is continuing these observations through its 
\textit{Synoptic Optical Long-term Investigations of the Sun} (SOLIS) facility. These different data sets can be combined into
a single disk-integrated \caii index time series that describes the average properties of the chromospheric 
emission over several solar cycles. 
We present such a \caii composite and discuss its correlation with
the new entirely revised sunspot number data series.
For this preliminary investigation, the scaling factor between pairs of time series
was determined assuming a simple linear model for the relationship between the monthly mean values during the duration
of overlapping observations.
\end{abstract}
%

\end{opening}

%
\section{Introduction}

\begin{figure} 
\centerline{\includegraphics[width=0.6\textwidth,clip=]{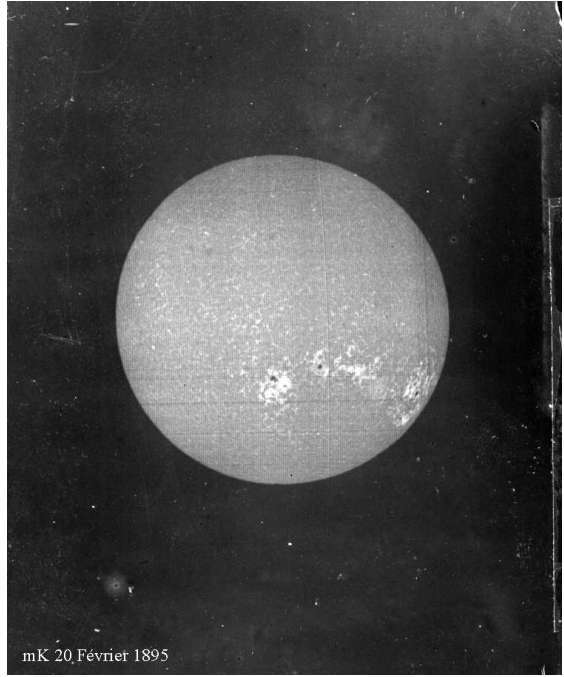}}
\caption{An early observation of the solar chromosphere in the \caii line. This image was taken on
February 20, 1895 and it is part of the Meudon archive of historical spectroheliograms available
at bass2000.obspm.fr/gallery2/main.php. The 
bright patches wisible on the disk are known as plages. They are associated with concentrations of magnetic fields and 
form a part of the network of bright emissions that characterize the chromosphere.
        }
 \label{history}
 \end{figure}

The first spectroheliograph was developed by George Ellery Hale in 1889 and
within a few years the first images of the Sun in the \caii line became available. Figure \ref{history} shows
an example of a very early observation taken with a spectroheliograph. The image is from the
Meudon archive of historical spectroheliograms, available at bass2000.obspm.fr/gallery2/main.php. 
Clearly visible in this image is the presence of brighter regions on the solar disk, known as
plages. However, it was only after the turn of the 20th century that regular observations in the \caii
line began at the Kodaikanal Observatory (India) and later, in 1915, at the Mount Wilson Observatory (USA).
Within the next decade, analogous programs began at several other observatories.                
Full-disk \caii observations started in 1917 at the National Solar Observatory of Japan (\exgr \opencite{2013JPhCS.440a2041H}), in
1919 at the Paris-Meudon Observatory in France (\exgr \opencite{1990A&A...227..577M}), in 1926 at the "Donati" solar tower telescope of the
Arcetri Astrophysical Observatory in Italy (\exgr \opencite{2009A&A...499..627E}), and in 1926 at the Astronomical Observatory of the Coimbra
University in Portugal (\exgr \opencite{2011CoSka..41...69G}).

Detailed studies of long-term variations in the solar activity cycle require observations taken on a continuous basis for 
several decades. At present, only four sets of direct observations of the solar atmosphere
are available for this purpose: 1) the visual sunspot counts, 
that retrace the last four centuries of solar activity; 2) the measurements in the ionized \caii line (393.37 nm); 
3)
H$\alpha$ observations; and 4) white-light measurements. Regular observations in \caii, H$\alpha$, and white-light
have been available since the early years of the twentieth century. Although all four sets of measurements provide important 
insight into the behavior of the 11-year solar cycle, the last three offer much broader diagnostic capabilities.
They can capture the complex hierarchical structures of the solar magnetic field and its evolution on a wide range of different 
spatial and temporal scales. In particular, \caii measurements in bright magnetic features 
can be used as a proxy for the line-of-sight unsigned magnetic flux density
(\exgr \opencite{1955ApJ...121..349B}; \opencite{1989ApJ...337..964S}; \opencite{2005MmSAI..76.1018O};
\opencite{2009A&A...497..273L}; \opencite{2016A&A...585A..40P}). 

Furthermore, observations and empirical models have shown that
solar irradiance varies on time scales from minutes to decades (\opencite{2004A&ARv..12..273F}; 
\opencite{2013ARA&A..51..311S}) 
and this variability is highly
correlated to area variations of plages and chromospheric network that can be observed in the \caii line 
(\exgr \opencite{2009SSRv..145..337D}; \opencite{2011CoSka..41...73E}; \opencite{2011JGRD..11620108F,2015ApJ...809..157F}
and references therein).
In particular, plages and chromospheric magnetic network play a significant role in modulating
the solar irradiance in the UV and EUV spectral bands which directly influence the Earth's 
atmosphere, from the stratosphere down to the troposphere
(\opencite{2006SSRv..125..331H}; \opencite{2009SoPh..255..229F}). 
Due to the Earth's atmospheric absorption it was not until long-duration measurements from space were available that changes in 
total and spectral solar irradiance were accurately measured. Those measurements began with the launch of the 
\textit{Nimbus 7} satellite in November 1978. However, \caii measurements can be used as a proxy to reconstruct the past history of solar irradiance over much longer
periods of time (\opencite{1998GeoRL..25.2909F}; \opencite{2003EOSTr..84..205F}).  

The purpose of this paper is twofold: 1) Create a 100+ years long, ongoing, disk-integrated \caii emission index time series
by combining both historical and modern data; 2) Investigate possible significant differences between this time series
and the recently released new sunspot number data. We anticipate that this new \caii time series may provide a useful tool for
a variety of studies related to the investigation of the long-term properties of the solar cycle. In the next section we
briefly discuss the properties of the three data sets used in creating the disk-integrated \caii emission composite, and how those
data were combined together into a single time series. In Section 3 we present the comparison between our new \caii composite
and the recently revised sunspot number data series. Our conclusions are highlighted in Section 4. 

\section{Disk-Integrated \caii Emission Indices}

For our analysis we consider data from three separate sets of measurements in the \caii line: 1) Data from
the Kodaikanal photographic
archive of spectroheliograms; 2) Data from integrated sunlight observations carried out at the National Solar Observatory
Sacramento Peak, and 3)
Data obtained with the \textit{Integrated Sunlight Spectrometer} operated by the National Solar Observatory in Arizona. 
These measurements and their corresponding \caii index time series,
have been described in great details in several previously published
studies. Here we only provide a brief summary of their properties.

\subsection{Kodaikanal Time series}

The archive at the Kodaikanal Observatory (KKL) of the Indian Institute of Astrophysics in Bangalore hosts the longest record of spectroheliograms
currently available.
After the Cambridge spectroheliograph was put into operation, in late 1904, there was some problem with the settings of the second slit at the correct wavelength. 
Although spectroheliograms were obtained in 1905 and 1906, the instrument did not perform to its potential until 1907, when regular observations began
(\opencite{2010ASSP...19...12H}).
The 70-micron exit slit of the spectroheliograph corresponds to a 0.5 \AA~ bandpass centered on the core of K3 at $\lambda$393.37 nm.
Spectroheliograms in the \caii line were obtained using the photographic emulsion until 2007. 
In particular, from the 1970s to late 2007 Scientia EM 23D56 emulsions from the AGFA company was used. After this stock was exhausted, in 2007, 
images in \caii were taken using Kodak film, but the results were not satisfactory and the program was terminated the same year. Observations
in \caii at Kodaikanal Observatory were also acquired using other instruments. For example,
beginning in 1997 observations
were obtained using a filter-based instrument (Daystar), with a 1.2\AA~ bandpass, and Photometrics CCD camera. After 2008 \caii measurements, 
along with broadband images of the Sun, were recorded on two ANDOR CCD cameras simultaneously using the TWIN telescope. 
While the TWIN telescope is still operational, due to problems with the CCD cameras, observations in Ca K line 
have not been obtained since October 2013.
Overall, the number of days available at Kodaikanal for making observations of the Sun had become less in the later part of the 20th century as compared to the first half of 20th century. 
Our investigation does not require the longest database available from this program of observations, but rather the most uniform and homogeneous one. For this
reason we used here data from a collection of 
more than 26 000 observations acquired between 1907 and 1999 that were digitized by \cite{2004IAUS..223..125M}. During this 
time interval
observations were taken with minimum changes in instrumental set-up, property of films, \textit{etc...}, providing a highly uniform
and homogenous set of images. This is the same data set investigated, for example, by \cite{2009SoPh..255..239T},
\opencite{2009ApJ...698.1000E} and \cite{2009SoPh..255..229F}.
The 8-bit depth and $\sim$2.0 arcseconds/pixel image scale of these images are sufficient to study plage areas with high 
contrast.
Recently, the same collection of \caii spectroheliograms has been re-digitized using a
4K$\times$4K CCD camera format with a pixel resolution of 0.86 arcseconds, and 16-bit readout 
(\opencite{2014SoPh..289..137P}). The
calibration of the new images includes flat-fielding, density to intensity conversion, and corrections for image geometry. 
About 80\% of those
images have been used in the study by \cite{2014SoPh..289..137P}, but due to some residual calibration issues the data have not 
yet been
released (J. Singh, private communication).

\begin{figure} 
 \centerline{\includegraphics[width=1.0\textwidth,clip=]{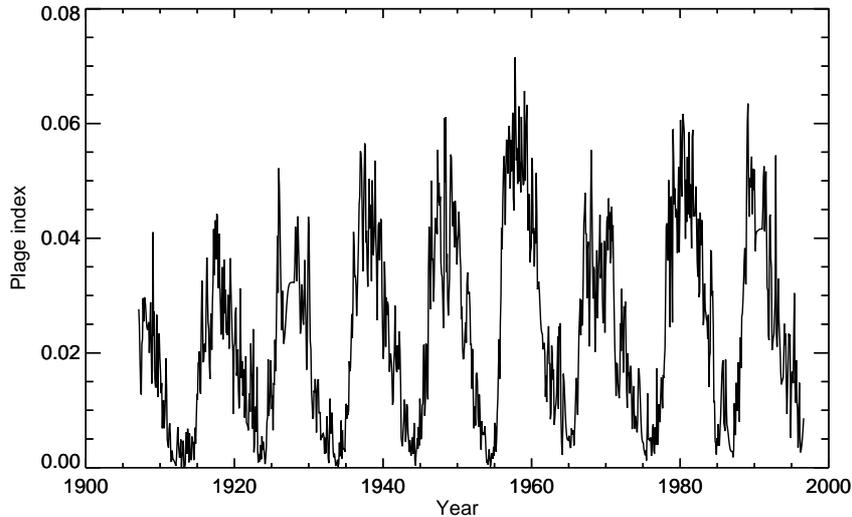}}
 \caption{Monthly averages of \caii plage index, in units of fraction of the visible solar hemisphere covered by plages and 
          active network derived from daily measurements taken at the Kodaikanal Observatory between 1907 and 1999.
  }
 \label{kkl}
 \end{figure}

In general,
the proper photometric calibration of photographic spectroheliograms
requires the presence of both the solar image and a series of stepwedge exposures on the plate. When this is the case, the 
transparency of the photographic plate material and the corresponding emulsion response curve can be determined using the
approach suggested by \cite{1968ApOpt...7.1513D}. Unfortunately, the lack of stepwedge exposures on a significant
portion of the KKL data makes difficult to consistently calibrate these images. In addition, changes in the image center 
across the field-of-view may produce variations in intensity and its gradient (vignetting) from one image to another. 
It should be noticed that these issues 
are common also in other similar historical data, such as the Mount Wilson archive of ionized \caii line spectroheliograms
(\opencite{2010SoPh..264...31B}). As discussed in \cite{2009ApJ...698.1000E}, properly addressing these issues is critical for the diagnostic
value of these measurements in long-term solar activity and irradiance studies. Different techniques have been proposed in the
past to mitigate artifacts affecting the quality of these data 
(\exgr \opencite{1998GeoRL..25.2909F}; \opencite{2005MmSAI..76..862L}; \opencite{2009SoPh..255..239T}; 
\opencite{2005MmSAI..76..941C}). 
In \cite{2009SoPh..255..239T} the same approach was used to calibrate \caii images from three different observatory: 
Kodaikanal Observatory,
Mount Wilson Observatory, and the National Solar Observatory at Sacramento Peak. A plage index was defined as
the fraction of the visible solar hemisphere covered by chromospheric plages and network elements at any given time
for all of these three sets of measurements.
For the first two sets of measurements, area values determined from the images were not compensated for foreshortening.
The \caii plage index time series derived from the Kodaikanal measurements is shown in Figure \ref{kkl}.

\subsection{NSO Sacramento Peak Time series}

Sun-as-a-star \caii line measurements have been taken with the
coelostat and horizontal Littrow spectrograph of the National Solar Observatory Evans facility at
Sacramento Peak (New Mexico, USA) since late 1976 (\opencite{1984ApJ...276..766K}). The instrument was designed to
have a very low level of scattered light by blocking the light reflected from the back of the spectrograph lens,
and from using the Littrow spectrograph in double-pass mode. 
The observed spectrum has a stepsize of 5.5 m\AA, and each \caii profile is flux-calibrated according to the
procedure described in \cite{1978ApJ...226..679W}. In short, the average flux computed within a 0.528 \AA-wide
window that is 1.187 \AA~ redward from the core of the \caii line is assumed to be equal to 16.2\% of the continuum value
near the \caii line. The equivalent width of a 1-\AA~ spectral band centered on the core of the \caii line is defined as
the \caii 1-\AA~ emission index (\opencite{1968ApJ...153..221W}). 
Because a similar program of observations (SOLIS/ISS)
began at Kitt Peak near the end of 2006, after reaching a sufficient overlap in time between the two series, 
the Sacramento Peak program was discontinued in November 2015.
Over 36 years of data from this program were recently analysed by \cite{2013ApJ...771...33S} to study the temporal variability of
seven different chromospheric parameters.

\subsection{SOLIS/ISS Time series}

The \textit{Integrated Sunlight Spectrometer} (ISS) is one of three instruments comprising
the \textit{Synoptic Optical Long-term Investigations of the Sun} (SOLIS, \opencite{2011SPIE.8148E...8B} and reference therein). 
In operation since late 2006, the ISS takes high spectral resolution (R $\cong$ 300,000) daily observations of the Sun-as-a-star 
in nine different spectral bands covering a large range of wavelengths (\opencite{2011SoPh..272..229B}). 
The observations taken in the \caii spectral line centered at 393.37 nm is
the longest continuous data in the SOLIS/ISS data set. 
The spectral band for the \caii line covers about 0.05 nm, and thus, does not include the continuum near the \caii line. 
To overcome this limitation, the observed line profiles are normalized  using intensities at two narrow bands situated 
in the blue (393.147-393.153 nm) and red (393.480-393.500 nm) wing of the \caii line (\opencite{2014AN....335...21P}). 
Mean intensities in these two bands are scaled to match intensities in the "reference" spectral  line profile taken by the 
NSO Fourier Transform Spectrometer (FTS, \opencite{2007assp.book.....W}). This normalization also helps to remove any 
residual linear gradients in intensity in the spectral direction. Variation in intensity over the activity cycle
at the location of these normalization bands is estimated to be around 1\%, which is much smaller than the variation
in the \caii line emission component.
For a  more detailed description of the SOLIS/ISS data reduction we refer the reader to \cite{2011SoPh..272..229B}.

 \begin{figure} 
 \centerline{\includegraphics[width=1.0\textwidth,clip=]{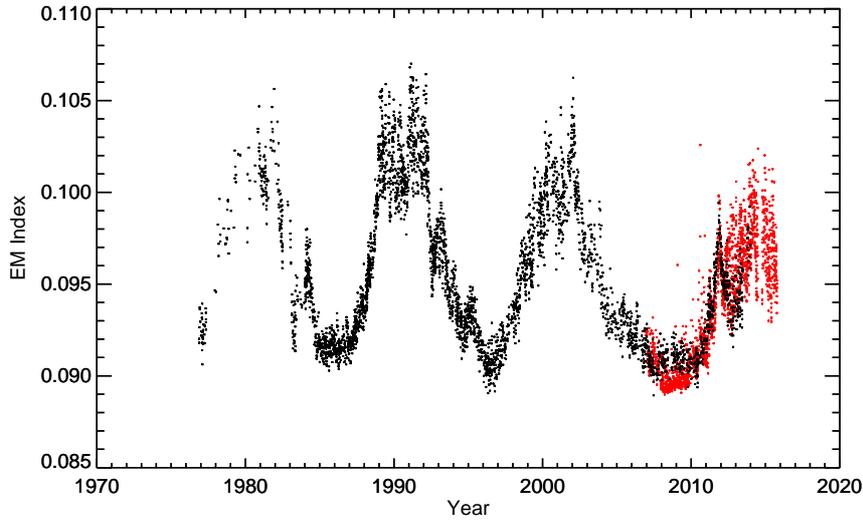}}
 \caption{Daily disk-integrated \caii 1-\AA~ emission (EM) index measurements from Sacramento Peak Observatory (black) and SOLIS/ISS (red) after
    adjusting the Sacramento Peak data.
  }
 \label{iss_sp}
 \end{figure}

\subsection{\caii Composite}

A composite of disk-integrated \caii measurements that dates back to 1907 can be produced by combining the three
time series described in Sections 2.1-2.3. The adopted Kodaikanal time series covers the 1907--1999 time interval, while data from the NSO/Sacramento
Peak time series were used up to the end of 2013. The ongoing SOLIS/ISS time series, that began December 2006, 
is used here as a reference for creating the composite.  
There is an overlap of about seven years (2007--2013) between the observations at Sacramento Peak and those taken by the SOLIS/ISS, which includes
the extended solar minimum of Cycles 23/24. This time-span is long enough to cross-calibrate the two sets of measurements.
Although the two instruments are very similar, and they measure the same quantities, they slightly
differ in sensitivity and noise level. This results in an offset between the two derived \caii emission index time series, with the Sacramento
Peak data being about 3\% lower on average than the ISS data. The correlation between the overlapping data sets is 0.88 after removing outliers greater than 
2.5 times the standard deviation and after removing yearly and monthly variations. 

\begin{figure} 
 \centerline{\includegraphics[width=1.0\textwidth,clip=]{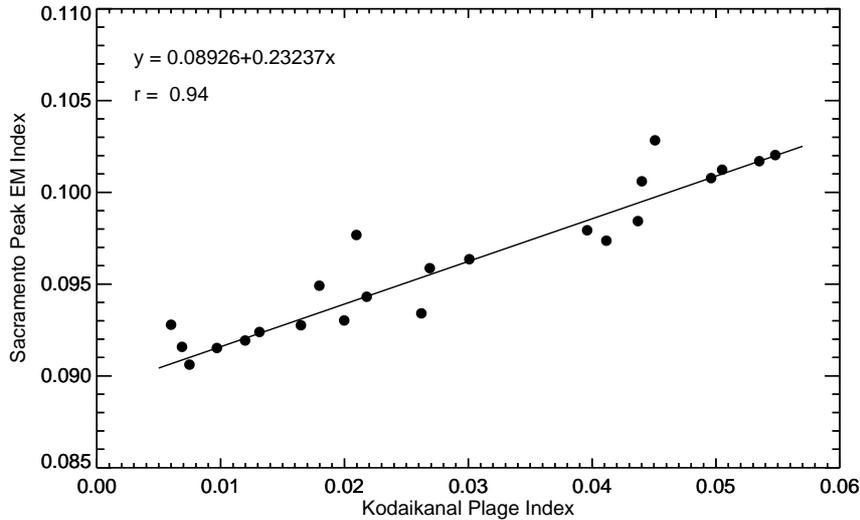}}
 \caption{Correlation between the Sacramento Peak emission (EM) index and the Kodaikanal plage index. Shown are the annual
mean values from both time series for the years 1978 to 1999, and the best linear fit to the data. 
The statistical significance of the correlation coefficient
$r$ is greater than 99.9\%. The resulting linear model, also shown in the plot,
allows the rescaling of the Kodaikanal plage index into an emission index.
  }
 \label{kd-sp_cal}
 \end{figure}

As a first step in creating the composite, we have adjusted the NSO/Sacramento Peak data to the SOLIS/ISS scale by 
multiplying the NSO/Sacramento Peak values by 1.032. This scaling factor was determined from the linear regression between the two
time series.  Figure \ref{iss_sp} shows the daily values of the two data sets,  after the correction was applied. A similar approach is then used
to rescale the Kodaikanal data to match the newly adjusted NSO/Sacramento Peak time series. Figure \ref{kd-sp_cal} shows the 
high correlation
between the annual mean values from both time series, for the 22 years (1978--1999) of overlapping observations. 
It should be pointed out that this high correlation is due
mainly to the strong 11-year cycle component presents in both time series. At shorter time scales the correlation is significantly
lower.

\begin{figure} 
 \centerline{\includegraphics[width=1.0\textwidth,clip=]{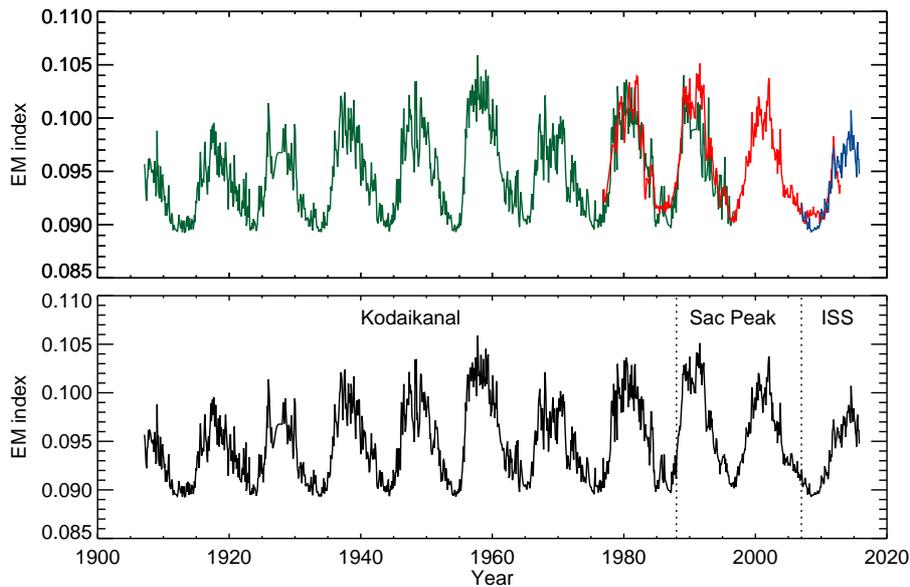}}
 \caption{
Top: Monthly disk-integrated \caii emission (EM) index derived from Kodaikanal (green), Sacramento Peak (red) and SOLIS/ISS observations 
(blue). Bottom:
Final composite obtained by merging the three time series together.
The vertical dotted lines indicate the transition from one time series to another.
  }
 \label{composite}
 \end{figure}

The paired data were fit to a linear model, $y = a + bx$, using a robust least absolute deviation method to mitigate the effect
of outliers. The parameters from this model, shown in Figure \ref{kd-sp_cal}, are then used to rescale the Kodaikanal plage index values
into an emission index. The top panel of Figure \ref{composite} shows the monthly mean values of the three data sets after the Kodaikanal
time series was linearly scaled. 
For times when observations were taken at two different observatories the comparison between those
measurements, at time scales longer than a month, shows some significant differences. 
We identified three periods when some discrepancy among the 
data is visible. A first period
of about two years, centered around 1992, where the Sacramento Peak data shows the presence of a more pronounced second maximum 
in Cycle 22 than indicated by the Kodaikanal data. The presence of this second maximum is also confirmed by other indicators of
solar activity, such as the newly released sunspot number time series discussed below and the 10.7 cm radio flux data (\exgr 
www.ngdc.noaa.gov/stp/solar/flux.html). 
A second period of about one year, from middle 1985 to middle 1986, shows a peculiar behavior in the Kodaikanal time series in the form of
a sudden rise in the plage index value visible in both Figures \ref{composite} and \ref{cak_sn}. 
A close inspection indicates that
daily calibrated measurements are mostly missing from May 15 to November 14, 1985 and for most of the second semester of 1986. In between, data are sparse and with values that suggest possible calibration issues. The monthly values used here are from the original 
plage index time series generated by \cite{2009SoPh..255..239T}, 
where missing values were linearly interpolated. This approach could 
potentially introduce some artifacts when, like in this case, several consecutive months of measurements are missing. This aspect
will be addressed in a more detailed study of the \caii time series planned for a near future.

Finally, during the extended minimum of Cycles 23/24 ($\sim$2008 -- 2010) the chromospheric emission
index derived from the Sacramento Peak measurements is about 1\% higher than the corresponding value derived from SOLIS/ISS. It also
shows an uncharacteristic flatness that seems to suggest possible issues with those data. On the other hand,
as demonstrated in the next section, the 
minimum in the emission index captured by the SOLIS/ISS data during that period agrees very well with the one from the sunspot number data.

Creating a composite requires to define the boundaries of each time series involved. Ideally, a new segment of the composite should
begin at the end of the previous one. However, due to some of the issues discussed above, we followed a slightly different
approach. The Kodaikanal data were used until the end of 1987, followed by the Sacramento Peak time series until December 31st, 2006.
The ongoing SOLIS/ISS data are then used starting January 1st, 2007. The resulting composite is plotted in the bottom panel of
Figure \ref{composite}. The continuity between the various time series is well preserved at the transition points, which are
indicated by the two vertical dotted lines. One can notice that at the sunspot minima, between Cycles 21--22 and Cycles 23--24, NSO/Sacramento Peak \caii data appear overestimating the level of the chromospheric activity as compared with KKL and SOLIS/ISS observations. Cancellation within ARs and active nests prevent the growth of plages, reducing the correlation at times of closely packed
active Sun.
These periods are not included in the composite time series shown in Figure \ref{composite} (bottom panel).
While the current proposed
series represents a significant first step, the reader should be cautioned that some remaining calibration issues may 
affect the quality of the proposed composite in some applications.

\section{Correlation with the Sunspot Number Time Series}

The sunspot number time series provides the longest direct record of the evolution of solar activity, covering a timespan of more than
400 years. This time series has been widely used in solar physics and climate studies. In recent years, however, several criticisms emerged
about the consistency and homogeneity of such a long database 
(\opencite{2002JAVSO..31...48H}; \opencite{1997JAVSO..26...40S}; \opencite{1997JAVSO..26...47S}). A major complication is due
to the fact that such a time series is constructed by combining observations taken from different observers, with different
instruments, and reduced using different processing techniques. A major effort was undertaken a few years
ago through a series of workshops dedicated to address most of those issues (\opencite{2013CEAB...37..401C}). 
The compendium of this endeavor is
summarized in a paper by \cite{2015sac..book...35C}, and by the release of a new revised sunspot number time series. 

 \begin{figure} 
 \centerline{\includegraphics[width=1.0\textwidth,clip=]{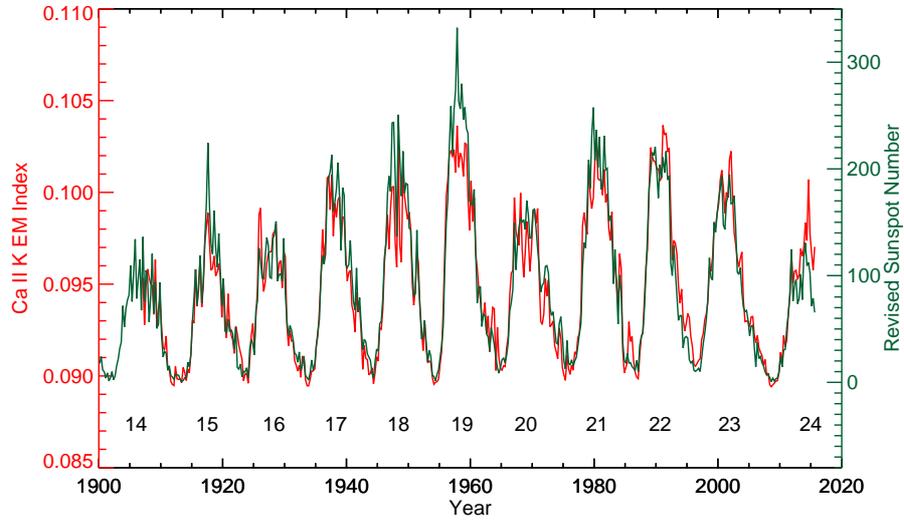}}
 \caption{
Quarterly averages of the composite \caii time series and the revised sunspot number. The numbers indicate the
different solar cycles.
  }
 \label{cak_sn}
 \end{figure}

The new time series is available from the Sunspot Index and Long-term Solar Observations (WDC-SILSO) website at
http://www.sidc.be/silso/datafiles (Source: WDC-SILSO, Royal Observatory of Belgium, Brussels). Some of the issues that
have been addressed in this new release include a correction for the Waldmeier sunspot weighting bias that was affecting the sunspot
number counting after $\sim$1947 (\opencite{2010ASPC..428..297S}), and the RGO-SOON scaling bias occurring after 1975 
(\opencite{2015sac..book...35C}).
Figure \ref{cak_sn} shows a comparison between our \caii composite and the sunspot number time series. 
They were plotted to have a common baseline, defined by their values during periods
of minimum of solar activity. The major discrepancy is during maximum of Cycle 19, where the minimum to maximum variation in the
sunspot number is significantly higher than observed in the chromospheric emission index derived from the Kodaikanal data. 
However, as pointed out by \cite{2009SoPh..255..229F} and \opencite{2009ApJ...698.1000E},
the relative amplitude in the Kodaikanal data during cycle 19
is significantly lower than in other similar time series such as the \caii plage index derived from the           
Mount Wilson photographic archive of spectroheliograms.
A possible explanation is that because the
Kodaikanal measurements were taken with a slightly wider spectral passband than those taken at the Mount Wilson Observatory they
refer to a deeper level in the solar atmosphere, where plage areas are smaller (\opencite{2007ASPC..368..533E}). 
\cite{2009SoPh..255..239T} also
indicated that, because of the narrower spectral bandpass, spot areas present in the Mount Wilson measurements are more difficult
to identify and separate from plages. 

 \begin{figure} 
 \centerline{\includegraphics[width=1.0\textwidth,clip=]{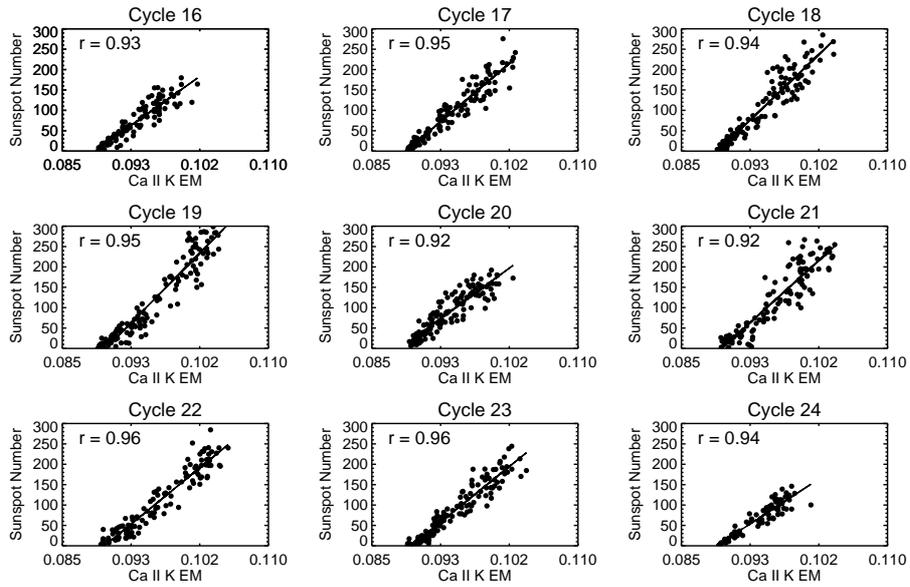}}
 \caption{Relationship between the monthly averaged disk-integrated \caii emission (EM) index and the revised sunspot number
for individual solar cycles. The Pearson's correlation coefficient for each cycle is indicated on the plots.
  }
 \label{cycle_month}
 \end{figure}

\begin{figure} 
 \centerline{\includegraphics[width=1.0\textwidth,clip=]{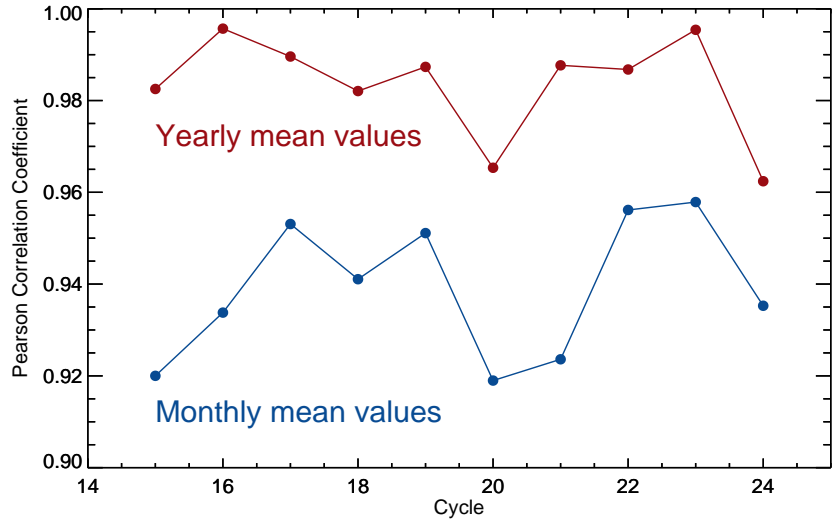}}
 \caption{Pearson's correlation coefficient between the \caii index and the revised sunspot number for individual solar cycles.
The correlation is computed using monthly and annual mean values from both time series.
  }
 \label{cycle_cor}
 \end{figure}

A more detailed analysis of the correlation between chromospheric emission index and sunspot counts is shown in
Figure \ref{cycle_month}. For each solar cycle we computed the Pearson's correlation coefficient between the monthly
mean values of both time series. Only months for which the sunspot number is greater than zero and \caii measurements 
are available were considered in computing these correlations. 
As for every ground-based synoptic programs of solar observations, continuity in the \caii data 
stream may be affected by weather conditions and/or instrumental issues. Consequently, individual monthly averages likely include a 
different number of observations and therefore the statistical significance of these values may vary.
The boundaries
for the different solar cycles were taken from en.wikipedia.org/wiki/List\_of\_solar\_cycles. Correlations with the sunspot number 
plotted in Figure \ref{cycle_month} exceed 0.92 for all the nine solar cycles, indicating an excellent agreement between the
two time series on time scales of a month or longer. It is also interesting to notice that the value of the slope derived from the
linear model greatly depends on the strength of the cycle: the stronger is the cycle and the larger is the value of this parameter.  
We have also repeated the same analysis using the annual mean values of
each time series. In this case the correlation significantly improves, reaching values as high as 0.99 for some of the solar cycles.
Figure \ref{cycle_cor} summarizes the results from both analyses. It should be noticed that current solar Cycle 24 has yet to be completed,
therefore reducing slightly the significance of the result in this case. There are some interesting facts that can be deduced from
a simple inspection of Figure \ref{cycle_cor}. The highest correlation between sunspot counts and chromospheric emission
index is reached during Cycle 23 for both the yearly and monthly mean values. This cycle is only the sixth strongest cycle among
those covered by the \caii observations described in this paper, but it shows a well defined double-maximum which is well captured
by both time series. Cycle 20 shows the lowest correlation value in both the annual and monthly mean value. This is mainly due to
the larger variability in the chromospheric emission index, compared to that of the sunspot count, during the phase of maximum
activity of that cycle (see Figure \ref{cak_sn}). Cycle 16 is an interesting case: it shows the highest correlation coefficient for
the annual mean values but a more modest value for the monthly averages compared to other cycles. Figure \ref{cak_sn} reveals that
Cycle 16 was characterized by a quite extended period of maximum of activity, of about five years, distinguished by significant
fluctuations in the sunspot counts over periods of few months. 
Unfortunately, the coverage of the Kodaikanal data during that period was quite poor, resulting
in a fragmented monthly series that required interpolation and consequently a reduced correlation with the corresponding sunspot
number series. The annual mean values are much less affected by the missing data, 
which explains the higher correlation coefficient in this case. 

\section{Conclusions}

In this paper we present a preliminary attempt to build an ongoing long-term \caii emission index time series by combining data from three
different sets of observations. Systematic observations in the \caii line began in 1907 at the Kodaikanal Observatory (India). 
While this program is still operational today, significant changes in the instrumentation after 2008 have affected the uniformity
of data time series. In particular, the use of a new filter with a much broader bandpass (1.2 \AA~ as compared to the original 0.5\AA) makes
very difficult to properly cross-calibrate the two subsets. In addition, calibrated public data are only available up to October 2007.
Although not
used in this investigation, it should be mentioned that a somewhat
similar program of observations was also established at the Mount Wilson Observatory between 1915 and 1985,
and at other observatories over different periods. Integrated sunlight 
measurements in the \caii line by the National Solar Observatory (USA) started at the Sacramento Peak Observatory
in the early 1970s, and continue today as a part of the 
\textit{Synoptic Optical Long-term Investigations of the Sun} (SOLIS) program.
Although quite different in nature, we show that the disk-integrated plage index derived from the Kodaikanal measurements
correlates well with 
the NSO/Sacramento Peak 1-\AA~ emission index at time scales as low as one year.
For more details about the comparison between the Kodaikanal measurements and other plage-index time series we refer the reader to the
papers by \cite{2009SoPh..255..239T}, \cite{2009ApJ...698.1000E}, \cite{2009SoPh..255..229F},
and \cite{2014SoPh..289..137P}. 
The significant overlap between pairs of those observations permits their cross-calibration and the ability to merge them
into a single long time series, as discussed in Section 2.4. 

This \caii composite covers more than 100 years of observations, or about 11 solar activity cycles. The cycle-to-cycle
variations in amplitude of our composite match very well those determined from the recent newly released sunspot number
time series, as shown in Figure \ref{cak_sn}. The only significant exception is for Cycle 19, the strongest cycle covered
by this study. As briefly explained in Section 3, a possible explanation is that Kodaikanal measurements during that period
were taken with a slightly wider spectral passband.
It should be pointed out, however, that the high correlation between
these two different time series is not necessary due to a very close relationship between \caii emission index and sunspot
number, but rather to the fact that both correlate well with the overall level of solar activity. In fact, this correlation breaks
down significantly at time scales of few days. When the analysis is preformed for each cycle separately, as shown in 
Figures \ref{cycle_month}-\ref{cycle_cor}, there is a clear linear relationship between these two quantities. Even at time scales of
about a month the Pearson's correlation coefficient remains quite high. 
Qualitatively, this could be understood in the following way: the brightness of the chromospheric plages in \caii line (and thus the sun-as-a-star emission index) correlate strongly with the magnetic flux. The sunspot number also correlates with the magnetic flux, but only with its portion represented by sunspots and pores. When the sunspots disappear, the magnetic flux of associated plages does not. Thus, on shorter time scales (days) sunspot number and emission index may not correlate very well with each other. 
On time scales larger than the typical lifetime of an active region (months), the monthly/yearly sunspot number will provide a good 
proxy for total magnetic flux in active regions (including plages) and thus, sunspot number and emission index will correlate well with each other.  
A previous study by \cite{2009SoPh..255..239T} has found
that the \caii plage index correlate quadratically with the sunspot area, therefore suggesting a non-linear relationship between sunspot
numbers and areas. This indicates that further investigations are required for better understanding the complexity of the solar cycle
of activity, and its various manifestations.

In a near future we aim to improve the quality of this composite, and make it publicly available to the solar community. For example,
the presence of daily gaps in the Kodaikanal time series could be mitigated by using a combination
of data from other databases, such as the Mount Wilson
archive of spectroheliograms. This will
improve the statistical significance of the monthly and annual average values. Also, a more detailed comparison of the daily measurements
obtained from different observatories should address some of the issues discussed in \cite{2009ApJ...698.1000E}. An improved \caii
composite will also offer a more powerful tool to understand better the physical significance of the sunspot counts. 

%
\begin{acks}
This work utilizes SOLIS/ISS data obtained by the NSO Integrated Synoptic Program (NISP), managed by the National
Solar Observatory, which is operated by the Association of Universities for Research in Astronomy (AURA), Inc.
under a cooperative agreement with the National Science Foundation. The sunspot number data are provided
by WDC-SILSO, Royal Observatory of Belgium, Brussels.
We thank the anonymous referee for insightful comments that led to a significant improvement of the overall 
quality of the article.
\end{acks}

%
%

\end{article} 
\end{document}